\pdfoutput=1

\documentclass[aps,prl,superscriptaddress,showpacs,reprint]{revtex4-1}
\usepackage{graphicx}
\usepackage{subfigure}
\usepackage{amsmath}
\usepackage{amssymb}
\usepackage{dsfont}

\usepackage{color} 

\usepackage{hyperref}
\hypersetup{
colorlinks=true,final=true,
        linkcolor=blue,
        citecolor=blue,
        filecolor=blue,
        urlcolor=blue,
        pdfauthor={L. Steffen},
        pdftitle={Realization of Quantum Teleportation with Solid State Qubits}
}

\begin{document}

\title{Comment on ``Vacuum Rabi Splitting in a Semiconductor Circuit QED System'' by Toida \textsl{et al.}, \textit{Phys. Rev. Lett.} \textbf{110}, 066802 – Published 6 February 2013}
\author{A.~Wallraff}
\author{A.~Stockklauser}
\author{T.~Ihn}
\affiliation{Department of Physics, ETH Zurich, CH-8093 Zurich, Switzerland}
\author{J.~R.~Petta}
\affiliation{Department of Physics, Princeton University, Princeton, New Jersey 08544, USA}
\author{A.~Blais}
\affiliation{D\'{e}partement de Physique, Universit\'{e} de Sherbrooke, Sherbrooke, Qu\'{e}bec, J1K 2R1, Canada}
\date{\today}
\maketitle


Toida \emph{et al.}~claim in their recent article~\cite{Toida2013} that they ``report a direct observation of vacuum Rabi splitting in a GaAs/AlGaAs double quantum dot (DQD) based charge qubit coupled with a superconducting coplanar waveguide (CPW) resonator'' \footnote{All text in colons is cited verbatim from Ref.~\cite{Toida2013}.}. They also claim that their work goes beyond results previously published in the literature \cite{Frey2012}, and in the process, neglect to cite other work \cite{Petersson2012a} in this area. In this comment, we challenge the main claims made in the paper \cite{Toida2013} and show that their results: a) do not provide any evidence of vacuum Rabi oscillations and b) do not provide any direct evidence of vacuum Rabi splitting.

In their paper Toida \textsl{et al.} state ``The distinct signature of the strong coherent quantum mechanical interaction between the two-level system (DQD) and the microwave photons (resonator) shows up when (i) ... : It manifests itself as two sharp parallel structures in a region between two paired charge triple points, as is shown in Fig. 3(b).'' However, the two sharp parallel structures in Fig.~3(b) of \cite{Toida2013} simply indicate the resonant interaction between the DQD and the resonator. This interaction is visible at detunings $\pm \epsilon$ corresponding to a crossing of the bare DQD transition frequency and the bare resonator frequency. The presence of these features is in no way indicative of a coherent quantum mechanical interaction.

More importantly, Toida \textsl{et al.}~claim ``The peak frequency of the spectra exhibits distinct anticrossings, as is shown in Fig. 4(b) [18], and the resonance line width increases significantly in the vicinity of the anticrossing points [Fig. 4(c)].'' Toida \textsl{et al.}~do indeed observe a small frequency shift of less than $2 \, \rm{MHz}$ due to the dispersive, i.e.~non-resonant, interaction between the DQD and the resonator in Fig.~3(b) of \cite{Toida2013}. However, a clear anti-crossing, allowing for a claim of the observation of strong coherent interaction of the vacuum-Rabi-type, is not observed as argued in more detail below. Importantly, the frequency range of the data displayed in Fig.~4(a) of \cite{Toida2013} is narrower than the suggested interaction rate $2 g/(2\pi)$, which does not even allow for the observation of the vacuum Rabi mode splitting in their data.


The key signature of strong coherent coupling of the vacuum Rabi type is the observation of a resonant mode-splitting with a pair of clearly identifiable distinct modes separated in frequency by $2 g/(2\pi)$, see for example Fig.~4b in Ref.~\cite{Wallraff2004} or Ref.~\cite{Haroche2006} and references therein. The line width of the \emph{two distinct modes} on resonance is $\Gamma = \gamma+\kappa/2$, with the resonator energy decay rate $\kappa$ and the DQD decoherence rate $\gamma = \gamma_1/2+\gamma_\phi$ determined by the DQD energy decay rate $\gamma_1$ and the pure dephasing rate $\gamma_\phi$ \footnote{We note that the expression for $\Gamma$ stated in Ref.~\cite{Toida2013} is incorrect.}. From their measurements Toida \textsl{et al.}~correctly determine $\kappa/(2\pi) = 8\,\rm{MHz}$. The key mistake in the analysis of the experimental data in \cite{Toida2013} is the following: They extract the line width of the data shown in Fig.~4(a) of \cite{Toida2013} (presumably using a fit to a single Lorenztian line) as displayed in Fig.~4c of \cite{Toida2013} and claim that the maximum observed value represents an accurate measure of $\Gamma$ on resonance. This is incorrect, as the presented expression for $\Gamma$
requires a resolved spectral measurement of the two vacuum Rabi modes to be applicable \cite{Haroche2006}. Toida \textsl{et al.}~mistakenly proceed to solve the expression of $\Gamma$ for the DQD decoherence rate $\gamma$. This procedure is incorrect and leads to a too small estimate of $\gamma/(2\pi)=12(25) \, \rm{MHz}$ \cite{Toida2013} eventually resulting in their unjustified claim of having observed the strong coupling limit.


\begin{figure}[!b]
\includegraphics[width=\columnwidth]{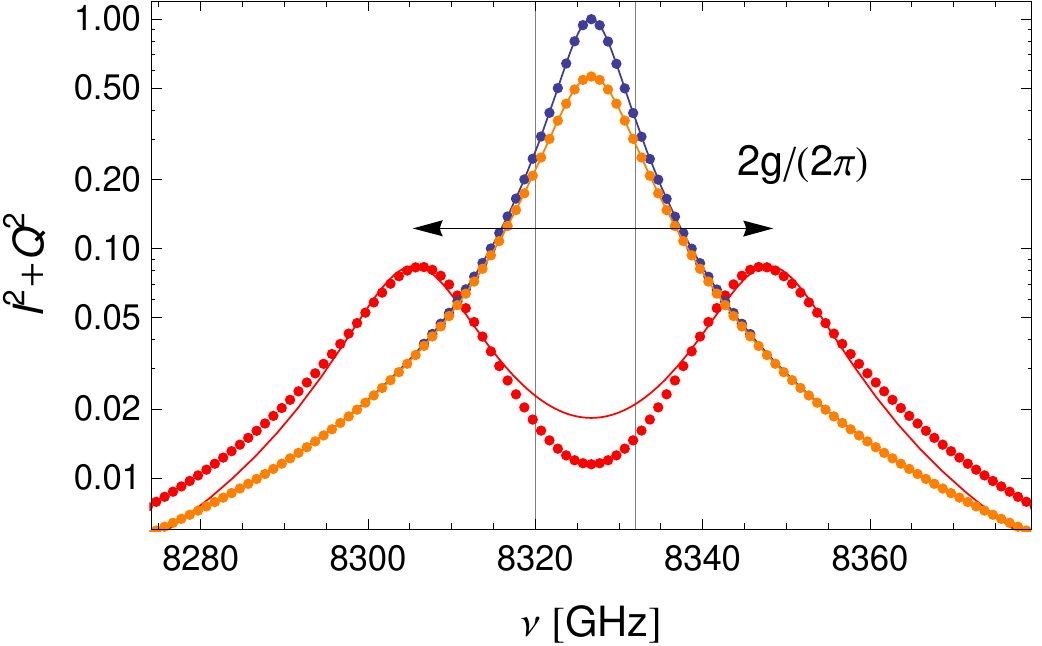}
\caption{Master equation simulation (dots) of transmitted power (on logarithmic scale) calculated as the sum of the squares of the in-phase (I) and out-of-phase (Q) quadrature of the transmitted microwave field as a function of drive frequency $\nu$. Solid lines are fits to Lorentzian line shapes. Red and orange data sets are with DQD on resonance with the microwave resonator for $(\gamma_1,\gamma_\phi,\gamma)/(2\pi)= (8,8,12)\,\rm{MHz}$ and $(200,200,300)\,\rm{MHz}$ respectively. For both data sets $\kappa = 8\,\rm{MHz}$ as indicated by the spectrum calculated with detuned DQD (blue data set) which we use to normalize the maximum power transmitted on resonance to unity.}
\label{fig:1}
\end{figure}

To demonstrate this fact we have simulated the expected transmission spectrum using a master equation simulation in the low photon number limit with the characteristic parameters extracted by Toida \textsl{et al.}, see Fig.~\ref{fig:1}. This simulation takes into account the energy relaxation and dephasing of both the resonator and the DQD. If the analysis by Toida \textsl{et al.}~was correct one would expect a clearly resolved vacuum Rabi mode splitting to be observable in their experiments as indicated by the red dots in Fig.~\ref{fig:1}. However, the authors do not present this essential data in their work.

\begin{figure}[!t]
\includegraphics[width=\columnwidth]{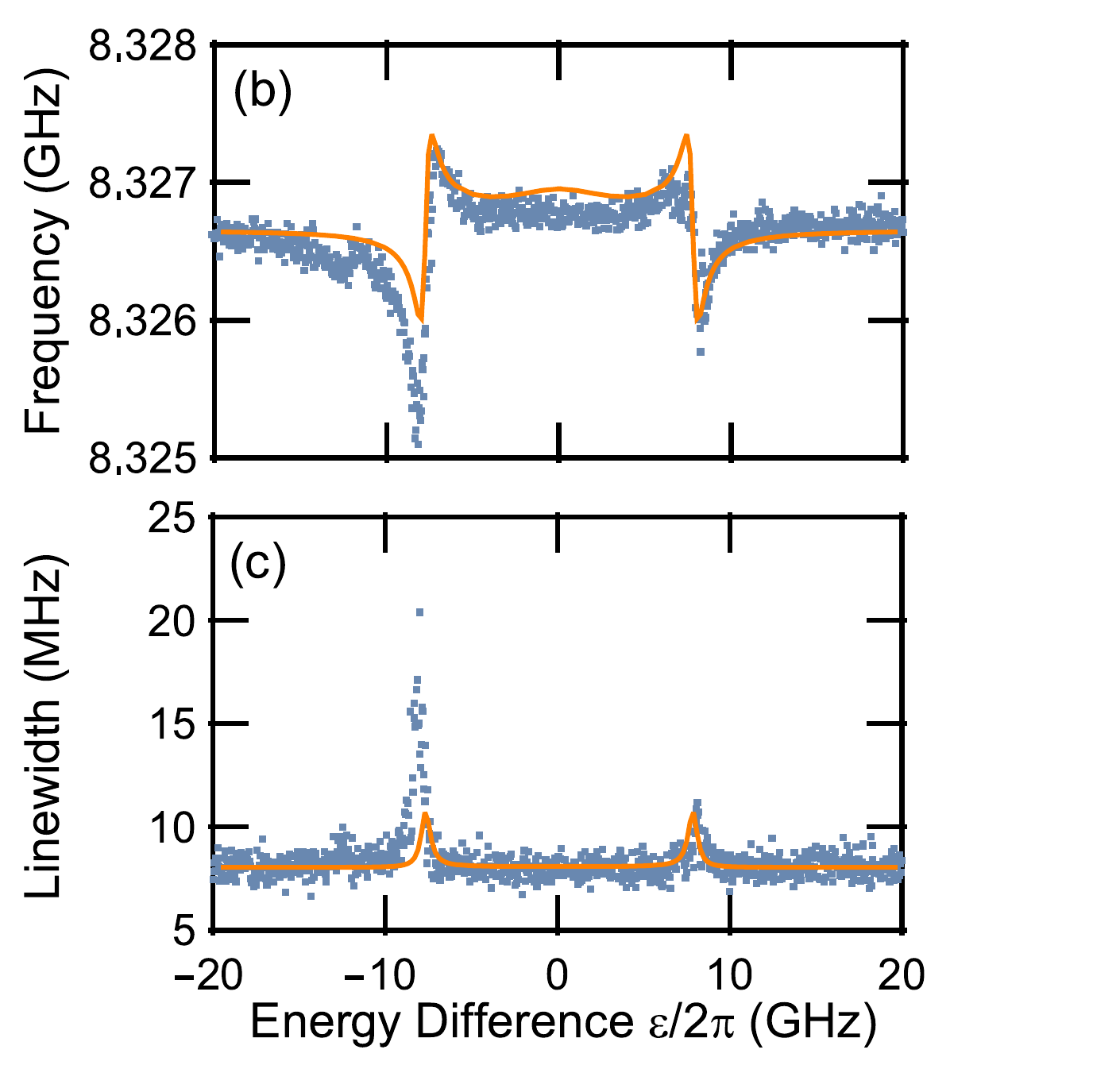}
\caption{Comparison of Fig.~4b and c of Toida \textsl{et al.} \cite{Toida2013} with our master equation simulation (orange line) 
using their parameters but the much larger DQD decoherence rates $\gamma_1/(2 \pi)=\gamma_\phi/(2 \pi)= \,200\,\rm{MHz}$.
}
\label{fig:2}
\end{figure}

Our analysis of the frequency shifts and line widths presented in Fig.~4(b,c) of \cite{Toida2013} using a master equation simulation \cite{Frey2012} including the counter-rotating terms of the Rabi-Hamiltonian results in values of $\gamma/{2\pi} = 300\,\rm{MHz}$ with $\gamma_1/(2\pi)=\gamma_\phi/(2\pi)=200\,\rm{MHz}$ rather than $\gamma/(2\pi)=12(25) \, \rm{MHz}$ as claimed in \cite{Toida2013}.
In our Fig.~\ref{fig:2} the simulated frequency shifts and line width extracted from a fit of the simulated spectra to a Lorentzian line shape (solid orange lines) are shown to be in reasonably good agreement with the data by Toida \textsl{et al.}

In addition we have performed analytical calculations of the resonator line width in a Jaynes-Cummings model, i.e.~using the rotating wave approximation, which are consistent with our numerical results.
In both the numerical and the analytical results the vacuum Rabi mode splitting is not observed, see simulated resonant spectrum (orange) calculated with these parameters in our Fig.~\ref{fig:1}. Also the line width extracted from both our simulations and analysis (orange line in Fig.~\ref{fig:2}c) is consistent with the data in Ref.~\cite{Toida2013}.


As a result, using the more realistic values for $\gamma/(2\pi) = 300 \, \rm{MHz}$ extracted from our analysis, the values for the number of Rabi flops $n_{\rm{Rabi}} = 0.07 \ll 1$, the critical photon number $n_0 = 112 \gg 1$ and the critical atom number $N_0 = 12 \gg 1$ all lead to the opposite conclusion that the strong coupling regime is \emph{not} reached in \cite{Toida2013}. Therefore, we are confident that the claim ``We hence conclude that the system is in a strong coupling regime with distinct vacuum Rabi oscillation [23].'' by Toida \textsl{et al.} is unjustified.
In conclusion the main claims presented in the paper \cite{Toida2013} are not supported by the experimental data.


We acknowledge discussions with Florian Marquardt and in addition comments on the manuscript from Julien Basset and Klaus Ensslin.

\end{document}